\documentclass[twocolumn,english,aps,prb]{revtex4-1}
\usepackage[T1]{fontenc}
\usepackage[latin9]{inputenc}
\usepackage{color}
\usepackage{amsmath}
\usepackage{amssymb}
\usepackage{graphicx}
\usepackage{esint}
\usepackage[normalem]{ulem}
\usepackage{babel}

\def\k{{\bf k}}

\def\red{\color{red}}

\definecolor{nb}{RGB}{63,0,255}
\begin{document}

\title{Effect of disorder on the competition between nematic and superconducting order in FeSe}

\author{ V. Mishra$^1$, and P. J. Hirschfeld$^{2}$}

\affiliation{~$^{1}$Computer Science and Mathematics Division, Oak Ridge National Laboratory,
Oak Ridge, TN  37831}
\affiliation{~$^{2}$Department of Physics, University of Florida, Gainesville,
FL 32611}

\date{\today}
\begin{abstract}
Crystalline FeSe is known to display strong nematic order below a weak
tetragonal-orthorhombic structural transition around $T_s\sim 90$K, and
a superconducting transition at $T_c\sim 9K$.  Recently, it was shown that
electron irradiation, which creates pointlike potential scattering defects,
has the surprising effect of enhancing $T_c$ while suppressing $T_s$.  Here
we discuss a possible scenario for such an effect, which postulates a 
competition between $s_\pm$ superconductivity and nematic order.  The transition
to the nematic state is modeled by a mean field theory of a $d$-wave Pomeranchuk
instability, together with a Cooper pairing interaction in both one- and 
multiband models.  The effect of nonmagnetic impurities on both orders is treated on
equal footing within the Born approximation.  We find evidence that disorder can 
indeed enhance $T_c$ while suppressing the competing nematic order, but only
in a multiband situation.  We discuss our results in the context of experimental 
data on FeSe crystals.
\end{abstract}
\maketitle
\section{Introduction.}
 Fe-based superconductors (FeSC) present a continuing challenge to the condensed matter
 physics community eight years after their discovery \cite{ChubukovHirschfeldPhysToday}.  
 Pairing is thought to be electronic in nature, and originate from  spin fluctuations,
 treated theoretically in various limits\cite{HirschfeldCRAS,Chubukov_bookchapter,Qimiao_Review2016}.  
 While many aspects of the superconducting state have been understood, the simplest material, 
 FeSe, has proven to be one of the most elusive.    It exhibits a tetragonal to orthorhombic
 structural phase transition at $T_s\sim 90\,\text{K}$, and displays very strong electronic 
 nematic behavior below this temperature ($T$), but never orders magnetically as do the 
 more familiar Fe pnictide systems.  While its critical temperature  $T_c\sim 9\,\text{K}$ is
 low for this class of materials, a sharp increase of  $T_c$ to about $40\,\text{K}$ is
 observed under a few GPa of  pressure, which simultaneously suppresses $T_s$ \cite{medvedev09,kothapalli16}. 
 Various intercalates of FeSe have $T_c$ of roughly this magnitude at ambient pressure as well, 
 and the highest $T_c$ of the entire class of Fe-based superconductors, $70\text{-}100\,\text{K}$,  
 is  achieved in monolayer films of FeSe grown on SrTiO$_3$ substrates.\cite{yan12,tan13,ge14} 
 Because of these intriguing hints of intrinsic higher-$T_c$ superconductivity,  efforts to
understand the properties of the basic bulk FeSe  have accelerated recently.

The origin of the nematic order in FeSe has itself been the subject of considerable
debate.  In general the origin of strong nematic tendencies in the FeSC have been
discussed in terms of a competition between fluctuations of  structural, orbital, 
and spin degrees of freedom\cite{Chubukov_Vavilov_Fernandes16}.  At first glance,
the lack of long-range magnetic order in the ambient pressure phase diagram seems
to suggest that the popular spin nematic scenario\cite{Christensen16} might not be
appropriate, and that orbital fluctuations might play a more important role\cite{bohmer13,Kontani15}.  
However, the confirmation of a long-range magnetic state under a modest 
pressure\cite{Bendele12,Terashima15} has lent support to other proposals that suggest
that the ground state at ambient pressure may be a quantum paramagnetic state\cite{FWang15,Glasbrenner15} 
or a state with long-range magnetic order of ``hidden'' quadrupolar type\cite{Yu15,Nevidomskyy16}. 
The very small Fermi surfaces in this system may also be important to prevent long-range ordering\cite{Chubukov_smallFS15}.

Recently, a new experiment by Teknowijoyo et al.\cite{Teknowijoyo16}  measuring $T_c$ and
superfluid density $\rho_s$ in FeSe has appeared that may provide an important clue to 
this puzzle.  The authors irradiated their sample with low-energy electrons, known to 
create homogeneously distributed Frenkel pairs of atomic vacancies and  interstitials 
without doping the system\cite{vanderbeek12}.  This seems to be the best way to create
pure potential disorder in these systems, where chemical substitutions have proven
difficult to interpret\cite{YWang13}.  In other Fe-based superconductors\cite{Prozorov_PH_irrad_Ba122,Shibauchi_PH_irrad_Ba122,Strehlow14}, 
$T_c$ has been strongly suppressed by disorder of this type, as compared 
to chemical substitution suppression rates.  In FeSe, on the other hand, 
Teknowijoyo et al.\cite{Teknowijoyo16} found an {\it increase} of $T_c$ 
upon irradiation, and concomitant {\it decrease} of $T_s$.  While these trends
are similar to the effect of hydrostatic pressure, the authors argued first that
irradiation tended to increase rather than decrease the lattice volume, and that
the magnitudes of the changes in lattice constants were in any case an order of
magnitude smaller than in pressure experiments.  They speculated that impurities
might introduce local pair strengthening effects due to the proximity of FeSe to
a magnetic transition\cite{Roemer12}, or that the competition of superconductivity
with nematic order might play a role.

 An intriguing paradigm to enhance $T_c$ by disorder for an $s_\pm$ state in the
 presence of competing stripelike ``$(\pi,0)$" magnetic order was offered by 
 Fernandes et al.\cite{TcEnhancement2012}, who pointed out that both interband and
 intraband impurity scattering processes would suppress magnetism, while only 
 interband processes would break superconducting pairs.  Whether  a competition
 of superconductivity with nematic order can similarly lead to a $T_c$ enhancement
 is not a priori obvious, given that nematic order is a $q=0$ distortion of 
 the Fermi surface. In addition, since the gap function in FeSe is known to 
 be highly anisotropic\cite{HirschfeldCRAS}, superconductivity will be 
 suppressed by both types of scattering processes, independent of any sign
 change over the Fermi surface.  From these perspectives, the $T_c$ enhancement
 mechanism of Ref. \onlinecite{TcEnhancement2012} should be irrelevant.

In this paper, we provide a concrete framework for the study of the effect
of impurities on the competition between nematic and superconducting order.
We first assume that that nematic order may be modeled by a $d$-wave 
Pomeranchuk instability competing with superconductivity.
Such an instability has been argued to qualitatively describe the nematic
tendencies of both cuprate \cite{Metzner05} and Fe-based superconductors\cite{DHLee_Pomeranchuk_2009,KhodasChubukov16}. 
We show here that the nematic order (Pomeranchuk instability) temperature, 
which we identify with $T_s$, is suppressed naturally by disorder, as 
found by earlier authors\cite{Schofield08}.  In addition, however, we 
find that $T_c$ can actually be enhanced when the nematic order is weakened.  
Whether $T_c$ is enhanced or suppressed by disorder turns out to depend in 
a nontrivial way on the interplay of the anisotropy of the superconducting 
pairing interaction with that  of the nematic distortion of the Fermi surface.

We now introduce the  mean field treatment of a $d$-wave Pomeranchuk distortion,
first within a single band model for pedagogical purposes, and then  
generalized to multiband Fermi surfaces appropriate to the FeSC.  Disorder
is then  treated in the Born approximation, and the instability temperatures
of both nematic and superconducting transitions calculated, for the different
model Fermi surfaces, and for different types of anisotropic $s$-wave 
superconductivity.  We present full phase diagrams of superconductivity 
and nematic order as functions of disorder and temperature, and discuss 
the comparison with experiment.

\section{1-band Model}
\subsection{Interplay of superconducting and nematic order}
 We wish to work with the simplest possible model capturing the competition of 
 superconductivity and nematic order, and allowing for the introduction of the
 effects of nonmagnetic disorder.  We begin therefore with a Hamiltonian describing
 fermions with an interaction which leads to a Pomeranchuk instability\cite{Metzner05},
${\cal H}={\cal H}_0+{\cal H}_{int},$ where ${\cal H}_0$ describes a noninteracting
parabolic band of electrons, ${\cal H}_0=\sum_{\k\sigma}(k^2/(2m)-\mu)c^\dagger_{\k\sigma}c_{\k\sigma}$, and 
\begin{eqnarray}
{\cal H}_{int}&=&\frac{V^{nem}}{4}\sum_{\k,\k'\sigma\sigma^\prime} (d_\k d_{\k'})\, c^\dagger_{\k\sigma} c_{\k\sigma} c^\dagger_{\k'\sigma'}c_{\k'\sigma'} \nonumber\\
&\simeq&  \sum_{\k\sigma}\Phi_{\k} c^\dagger_{\k\sigma}c_{\k\sigma},
\end{eqnarray}
where $\mu$ is the chemical potential, $d_\k=\cos 2 \phi$, and $\phi$ is the angle around the Fermi surface.
In the second line we have made a mean field approximation\cite{Metzner05} and
defined $\Phi_\k\equiv \Phi_0 d_\k =d_{\k}\sum_{\k'\sigma}\langle d_{\k'} c^\dagger_{\k'\sigma} d_{\k'\sigma} \rangle$, 
which leads to a self-consistency equation for the nematic order parameter 
\begin{eqnarray}
\Phi_\k &=&  -T d_\k \sum_{\omega_n,\k'}  \frac{ V^{nem} d_{\k'} \left( i\omega_n - \xi_{\k'}-\Phi_{\k'} \right)}{\omega_n^2+\left( \xi_{\k'}+\Phi_{\k'} \right)^2}. \label{Eq:Phi1band}
\end{eqnarray}
The nematic transition temperature when $\Phi_\k\rightarrow 0$,
$T_{nem}$, is then given by
 \begin{eqnarray}
 T_{nem}= \frac{\mu}{2 \tanh^{-1}\left(4 \lambda^{-1}_{nem}\right)},
 \label{Eq:Tnem1bnd}
\end{eqnarray}
where   $\lambda_{nem}=m V^{nem}/2\pi$. 
It is worth noting that the
nematic instability is of the Stoner type: unless $V^{nem}$
is larger than a threshold value, long range nematic order is not possible.

\begin{figure}[h]
\includegraphics[width= 0.8\columnwidth]{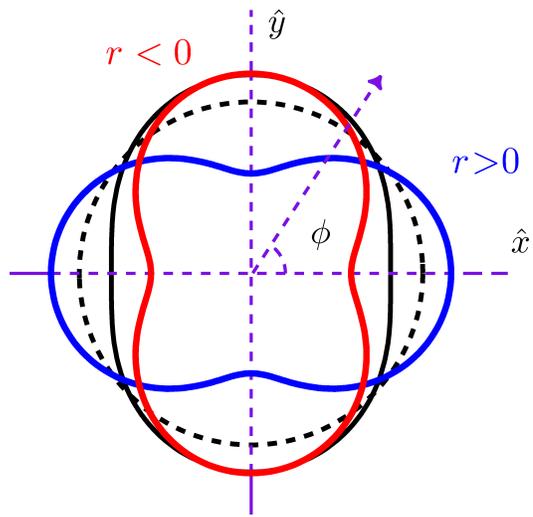}
\caption{Schematic Fermi surface at T=$T_{nem}$ (dashed black curve) and $T=T_{nem}/2$ 
(solid black), with momentum angle $\phi$ indicated. Superconducting 
gap with anisotropy given by $r>0$ (blue) and same with $r<0$ (red). }
\label{fig:FS1band}
\end{figure}

Next, we study superconductivity in the nematic phase, and show
that the interplay of nematicity and superconductivity depends sensitively
on the structure of the superconducting pairing.    We consider a simple
model of a $C_2$-symmetric gap, $\Delta_\k=\Delta_0 (1 +r \cos 2\phi)/\sqrt{1+r^2/2}\equiv \Delta_0{\cal Y}(\phi)$, 
where ${\cal Y}(\phi)$ is normalized to 1 over the high temperature  Fermi surface. 

To study the interplay of such states with nematic order, we introduce
the anisotropic pair potential $V_{\k\k'} = V^{sc} {\cal Y}(\phi){\cal Y}(\phi')$ on
the Fermi surface, and for the moment assume it  to be independent of nematic changes
in the Fermi surface itself.  The linearized gap equation in the presence of 
nematic order, $\Phi_0>0$ may then be expressed as 
\begin{eqnarray}
1&=& 2\pi T_{c0} \lambda_{sc} \sum_{\omega_n>0}^{\omega_c} \int_0^{2\pi}\frac{d\phi}{2\pi}\mathcal{Y}^2(\phi) \mathbf{P}(\phi,\omega_n), \\ \label{eq:GPE}
\mathbf{P}(\phi,\omega_n)&=& \frac{1}{\omega_n}\left[\frac{1}{2}+\frac{1}{\pi}\tan^{-1}\left(\frac{\mu-\Phi_\k}{\omega_n}\right)\right]. \label{eq:P0}
\end{eqnarray}
Here the pair bubble $\mathbf{P}(\phi,\omega_n)$ makes the maximum
contribution along the long direction ($\phi=\pi/2$) of the deformed Fermi
surface. If the minimum of the absolute value of gap function  is also along
this direction (\textit{i.e.} $r>0$), then the deformed Fermi surface will suppress such gap states.

We can approximate the effect of nematic order on $T_c$ 
in the limit of weak deformation of the Fermi surface ($\Phi_0 \ll \mu$),
\begin{eqnarray}
 \log \left( \frac{T_{c0}^{nem}}{T_{c0}} \right) \simeq -\frac{\pi r}{2+ r^2} \left( \frac{\Phi_{0}}{2\mu} \right),
 \label{eq:nemTc2}
\end{eqnarray}
here $T_{c0}^{nem}$ ($T_{c0}$) is the superconducting transition temperature
in the nematic (tetragonal) phase with no disorder.
As expected from the structure of ${\bf P}$, the critical temperature $T_c$ depends
on the structure of the pairing through the model anisotropy parameter $r$, and can
apparently increase or decrease depending on how the anisotropy is oriented with
respect to the distorted $C_2$ Fermi surface.   
When the Fermi surface gets stretched along the direction of gap  minima ($r>0$, Fig. 1) $T_c$ is suppressed.
In contrast,  gap minima oriented  perpendicularly to the stretched Fermi 
surface ($r<0$, Fig. 1) lead to an increase of  $T_c$, since ${\bf P}(\phi)$ 
and $\Delta(\phi)$ have maxima on the same regions of the Fermi surface. This is similar
to superconductivity in the presence of spin-density wave order, where gap 
maxima away from the nesting hot spots is favorable\cite{VMSCSDW}.

We note the important role of particle-hole asymmetry for this problem.  
As seen explicitly from Eq. (\ref{eq:P0}),  in the limit of perfect 
particle-hole symmetry ($\mu\rightarrow \infty$), the coupling between
superconducting and nematic order vanishes.  Furthermore, the sign of 
the nematic distortion depends on the sign of $\mu$: thus if one replaces
the electron band assumed above with a hole band, the effect will be  
identical, except that $T_c$ will now be suppressed for $r<0$.  The relative
orientation of the gap states described above remains unchanged, however: $T_c$ 
is suppressed relative to $T_{c0}$ if gap minima are along the stretched direction. 

\begin{figure}[h]
 \includegraphics[width= 1\columnwidth]{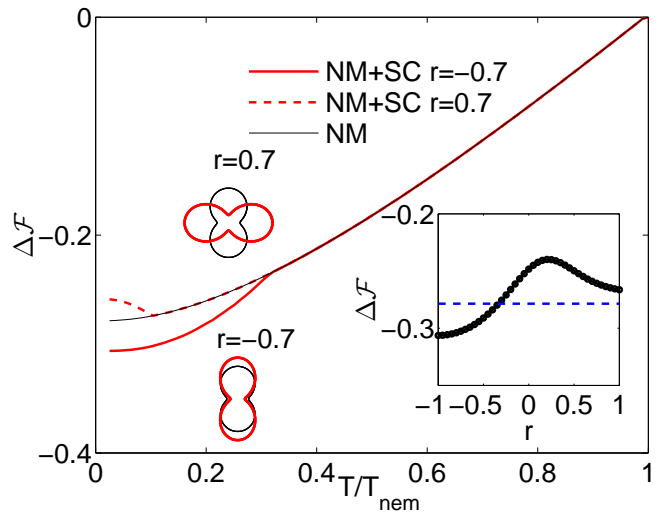}
\caption{{Competition of superconductivity (SC) and nematic order (NM).  
Total free energy relative to normal metallic state $\Delta {\cal F}$ vs. $T$ 
for pure nematic state and  two cases of superconducting and nematic coexistence,
with  superconducting anisotropy $r=\pm 0.7$.  Insert: $\Delta {\cal F}$ at $T=0.005\mu$  
as a function of superconducting anisotropy parameter $r$. Gap structures (red) 
are shown with respect to the deformed Fermi surface (black).}}
\label{fig:stability_1band}
\end{figure}

These discussions assume that the nematic/superconducting coexistence solutions
discussed above are ground states for the given parameters.  However, an 
examination of the free energy (see Appendix) shows that those superconducting 
states with gap minima along the nematic elongation of the Fermi surface are 
higher energy  than the homogeneous nematic state, and are therefore 
unstable, as illustrated in Fig. \ref{fig:stability_1band}.

\begin{figure}[h]
\includegraphics[width= 1\columnwidth]{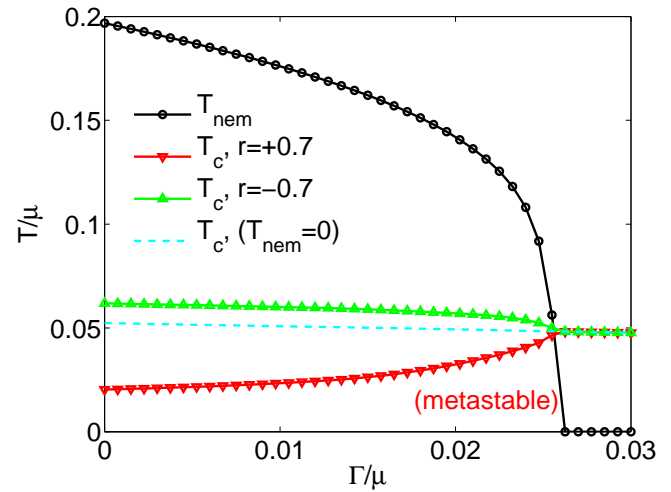}
\caption{ Nematic transition temperature as a function of disorder 
scattering rate (circles) $\Gamma=n_{imp}V_{imp}^2(m/2\pi)$, along 
with $T_c$ in the nematic phase for two different orientation of the 
gap function (triangles), normalized to chemical potential $\mu$. 
Variation of $T_c$ with disorder in the absence of the nematic 
order is shown by a dashed line.  }
\label{fig:Tnem1bnd}
\end{figure}
\subsection{Effect of disorder}
 We first determine the nematic order parameter in the presence 
 of nonmagnetic disorder above $T_c$ by defining a self-energy within the Born approximation,
\begin{equation}
\Sigma_{nem}(\omega_n) = n_{imp}|V_{imp}|^2 \sum_{\k^\prime} { G}(\k',\omega_n),\label{Eq:Sigma_nem_1band}
\end{equation} 
where  $n_{imp}$ is the concentration of impurities with potential $V_{imp}$, and $G(\k,\omega) =(i\tilde\omega -\xi_\k - \Phi_k +i0^+)^{-1}$
 at $T_{nem}$, and $i\tilde\omega = i\omega -\Sigma_{nem}(\omega)$.   This 
renormalizes the single-particle energy and influences the nematic order through 
its self-consistency equation.  Due to the $d$-wave Pomeranchuk form, 
the $\Phi_\k$ itself is not renormalized in this approximation.  

To study superconductivity within the same framework, we introduce the disorder
self-energy in Nambu space in the same approximation,
\begin{eqnarray}
 {\hat\Sigma}(\omega_n) &=& n_{imp} |V_{imp}|^2 \sum_{\k^\prime} {\hat G}(\k',\omega_n)\equiv \sum_{\alpha=0,3}\Sigma_\alpha \tau_\alpha,\label{Eq:Sigma_SC_1band}
\end{eqnarray}
where $\hat G$ is the Nambu Green's function
\begin{equation}
 \hat G(\k,\omega_n) = -\frac{i\tilde\omega_n \tau_0 + (\xi_\k +\Phi_\k)\tau_3 + {\tilde\Delta}_\k \tau_1}{\tilde\omega_n^2 + (\xi_\k+\Phi_\k)^2 + {\tilde\Delta}_\k^2}.
\end{equation}
Here the renormalized order parameter has the form $\tilde\Delta_\k \equiv \tilde \Delta_{iso} + \Delta_{ani} d_\k$, and we have defined
\begin{eqnarray}
 i\tilde{\omega}_{n} &=& i\omega_n -\Sigma_0(\omega_n), \label{Eq:wtild} \\
 {\tilde{\Delta}}_{iso} &=& \Delta_{0}/\sqrt{1+r^2/2} + \Sigma_1(\omega_n) , \label{Eq:deltild}\\
 \Delta_{ani}&=& \Delta_0 r/\sqrt{1+r^2/2}.
 \label{Eq:deltild_ani}
\end{eqnarray}
Note that the nematic order parameter $\Phi_\k$ and  anisotropic gap component $\Delta_{ani}$ are unrenormalized by disorder.

At $T_c$, the self-consistency expressions for the amplitude of the $d$-wave nematic
order parameter $\Phi_0$ and the superconducting order parameter $\Delta_0$ are given by
\begin{eqnarray}
\Phi_0 &=&  -T \sum_{\omega_n,\k}  \frac{V^{nem} d_{\k} \left( i\tilde\omega_n - \xi_\k-\Phi_\k  \right)}{\tilde\omega_n^2+\left( \xi_\k+\Phi_\k \right)^2}, \label{Eq:Phi1band}\\ 
  \Delta_0 &=& -T \sum_{\omega_n,\k} \frac{ V^{sc}  {\cal Y}(\phi) \tilde{\Delta}_\k}{\tilde{\omega}^2+(\xi_\k + \Phi_\k)^2} 
  \label{Eq:Tc1band}
\end{eqnarray}

Fig. \ref{fig:Tnem1bnd} now shows the result of a numerical evaluation 
of Eqs. (\ref{Eq:Phi1band})-(\ref{Eq:Tc1band}).  As expected, disorder 
gradually suppresses the nematic order as also found  in Ref. \onlinecite{Schofield08}. 
From the inset in Fig. \ref{fig:stability_1band}, we can anticipate that the $r>0$ case, 
which shows no competition, should
lead to a $T_c$ suppression with disorder, as indeed shown in Fig.  \ref{fig:Tnem1bnd}.   
For a small suppression of $T_{nem}$, $T_c$ appears to increase when $r>0$, i.e. when 
the gap minima are along the elongated direction of Fermi surface. 
 However, as shown in Fig. \ref{fig:stability_1band}, such a pairing state is a metastable state. 
Thus disorder-induced $T_c$ enhancement appears to be disfavored in this one band model.
 
\begin{figure}[h]
\includegraphics[width=0.38 \columnwidth]{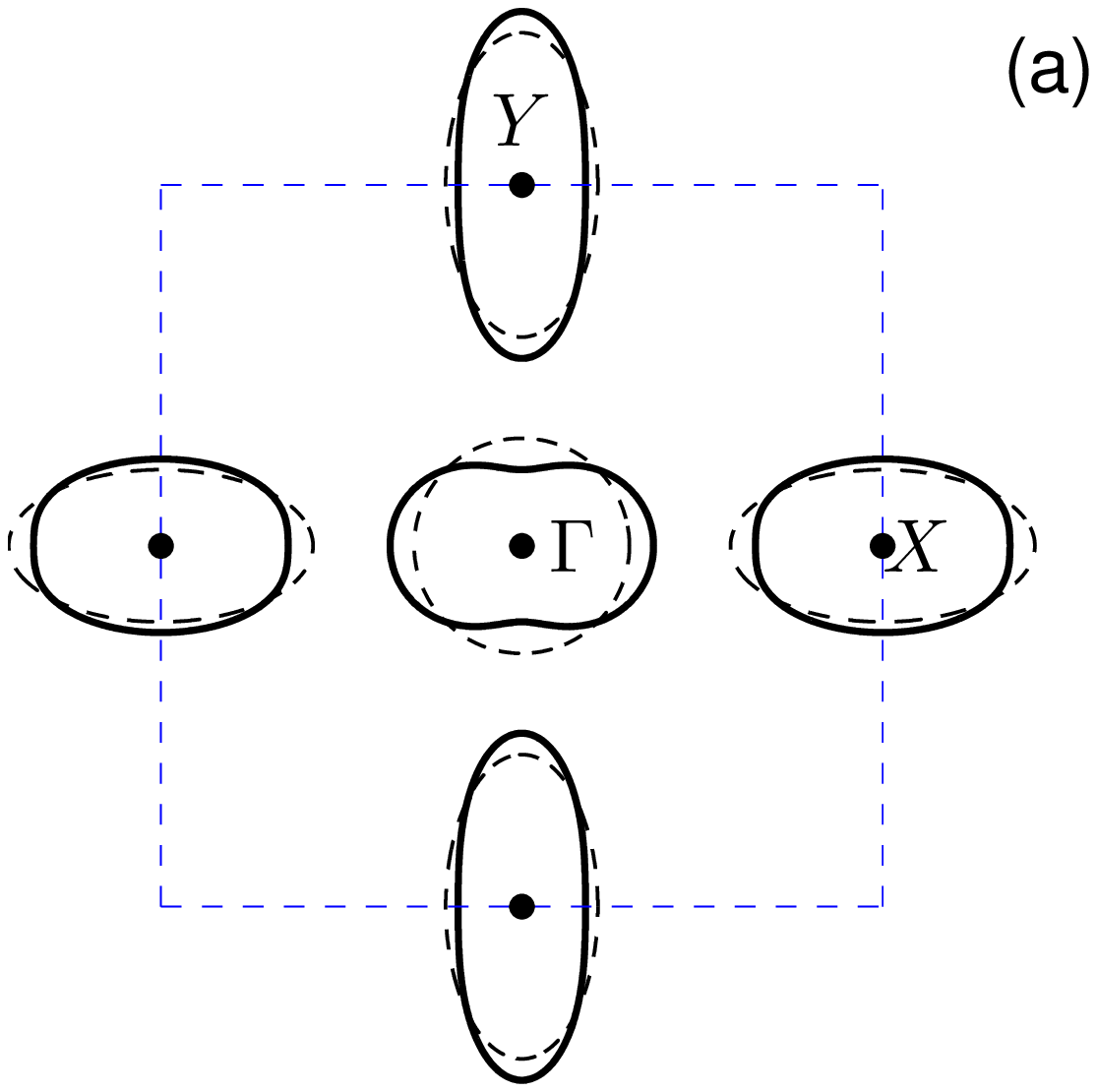}
\includegraphics[width=0.38 \columnwidth]{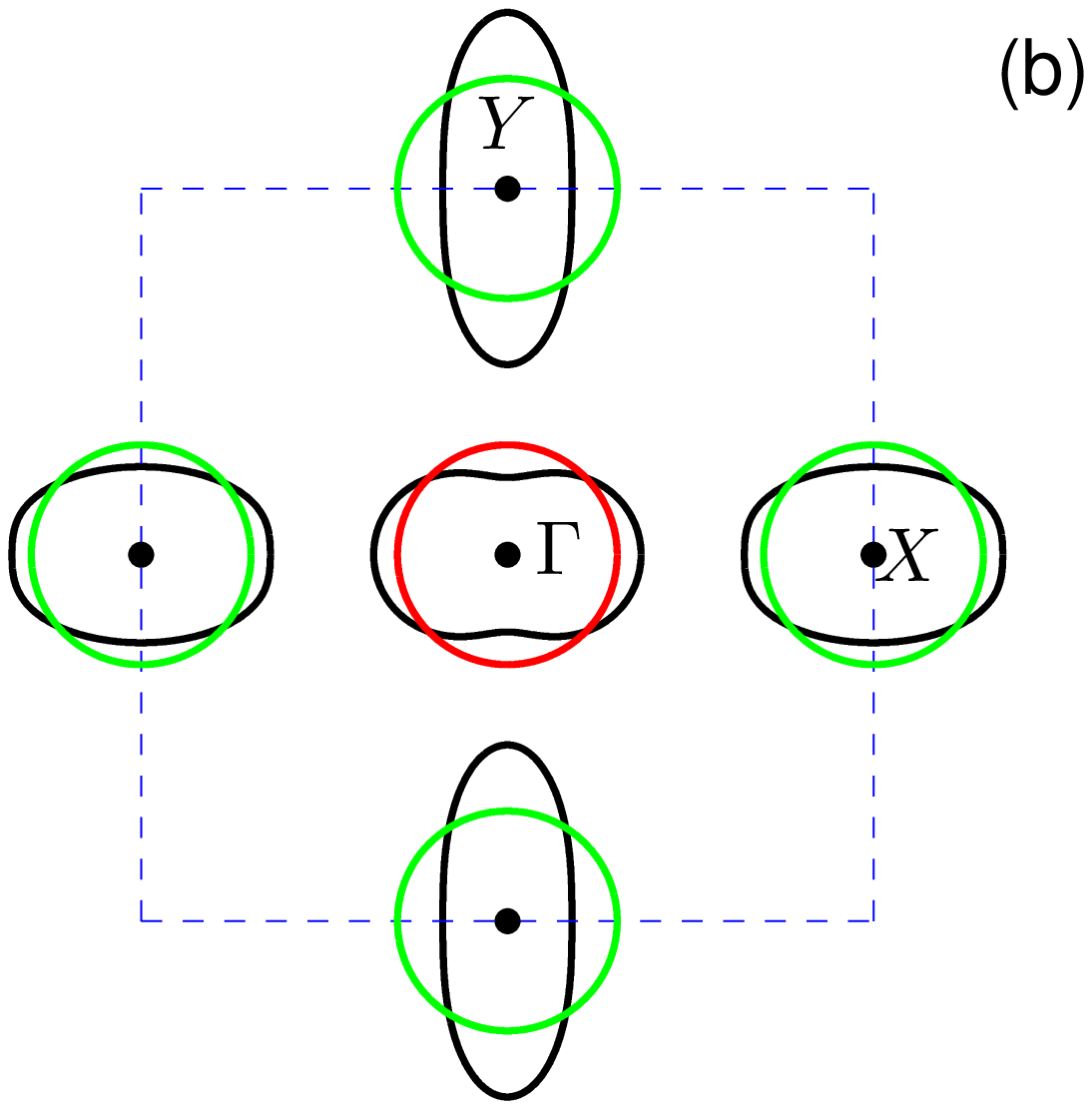}
\includegraphics[width=0.38 \columnwidth]{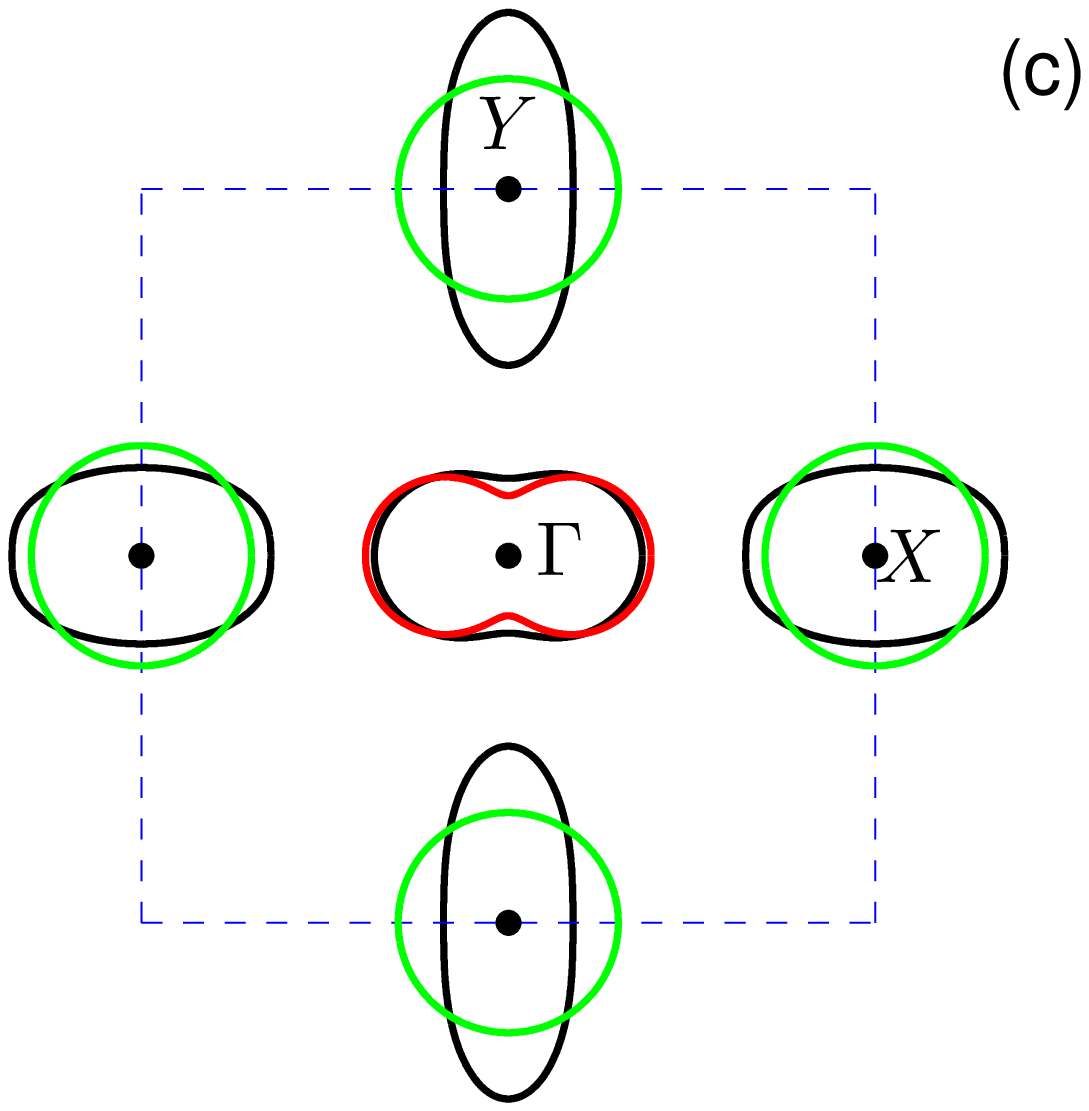}
\includegraphics[width=0.38 \columnwidth]{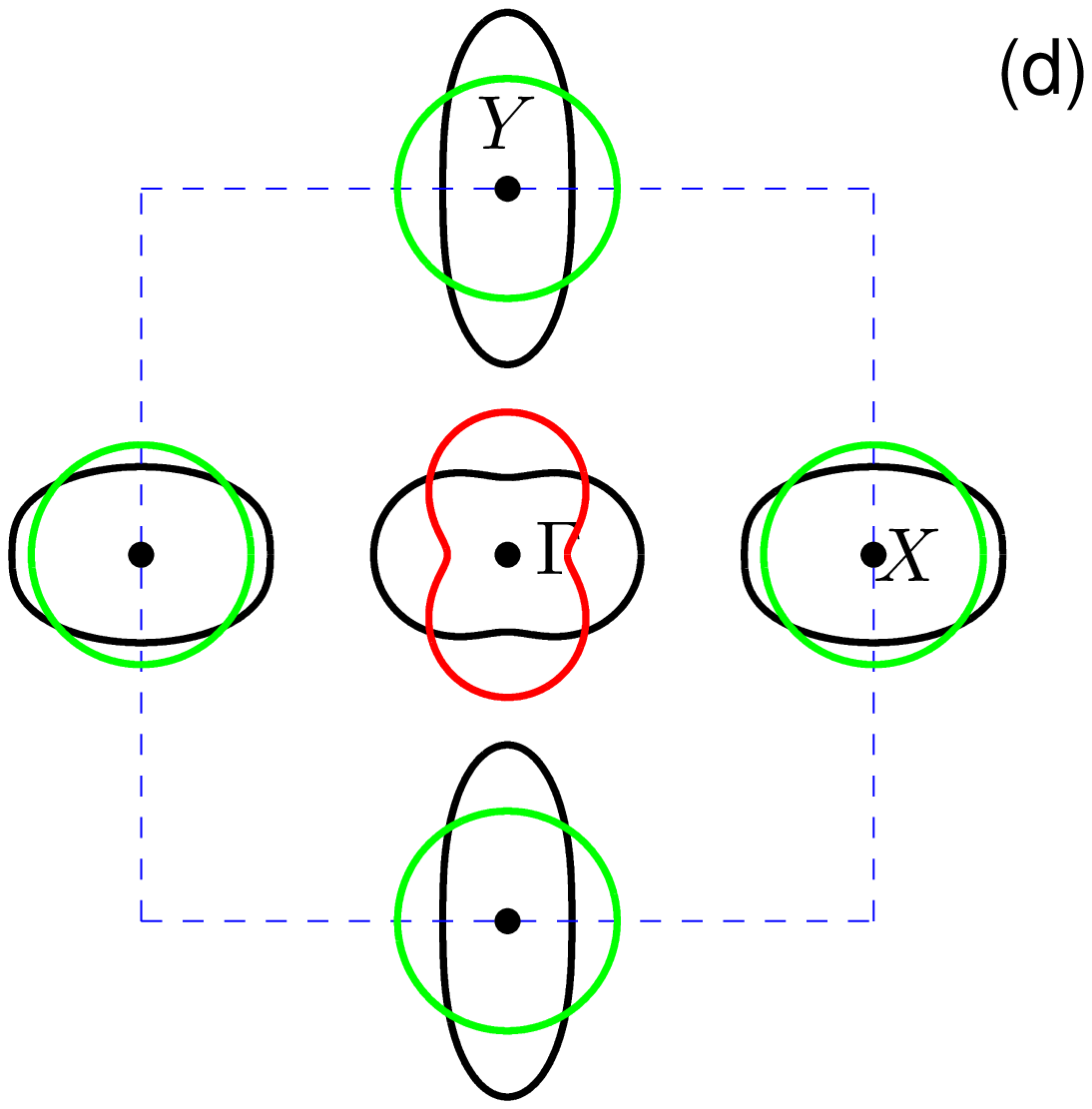}
\includegraphics[width=0.38 \columnwidth]{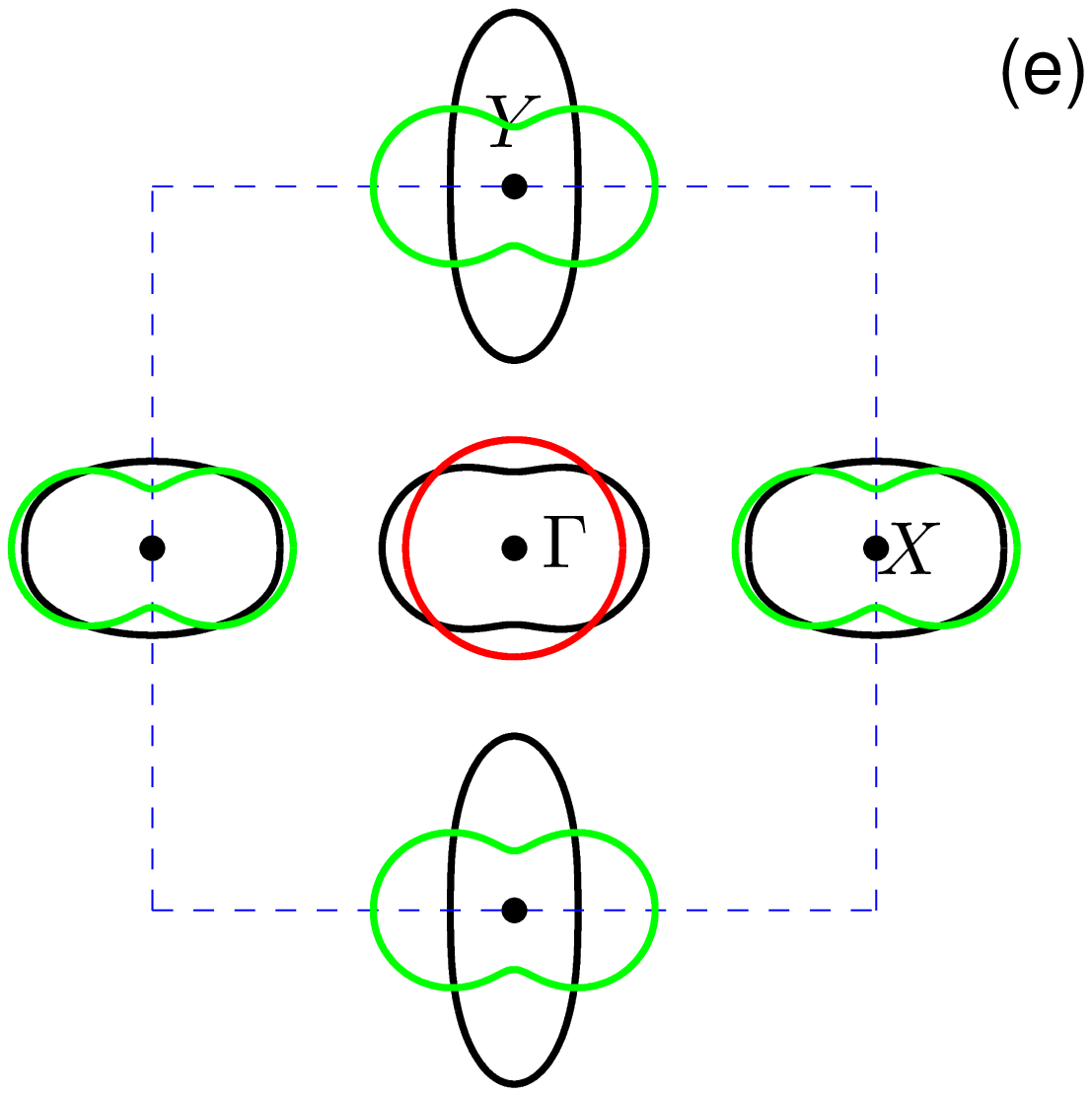}
\includegraphics[width=0.38 \columnwidth]{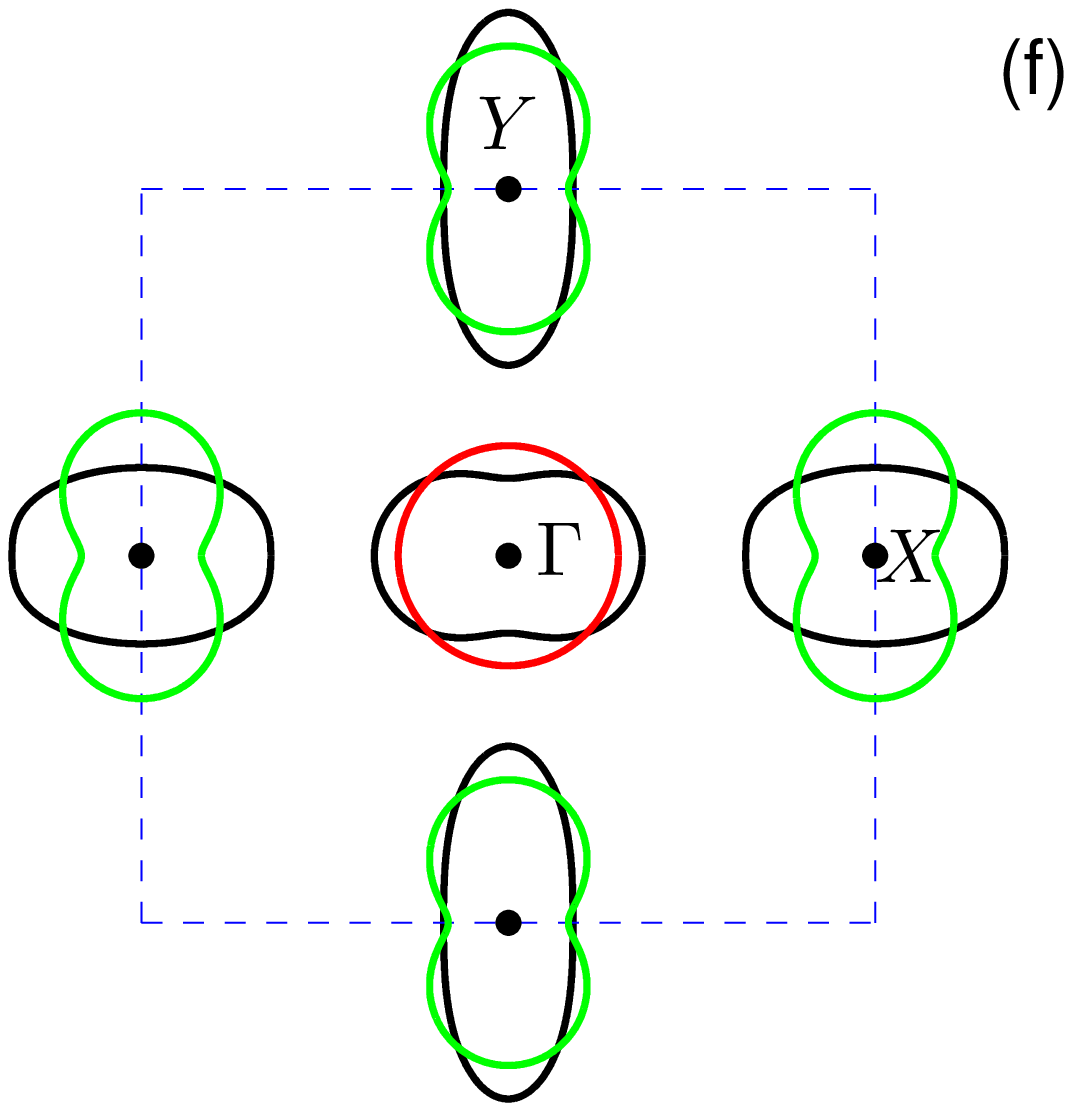}
\caption{(a) Fermi surfaces for the three band model shown in the paramagnetic phase (dashed line)
and in the nematic phase (solid line) using the 1-Fe zone representation. (b) isotropic pair states on all bands with opposite signs (illustrated with red and green ); (c) isotropic electronic band gaps and hole band gap $\sim 1+ r_h \cos 2 \phi$, $r_h >0$; (d) same but with $r_h<0$; (e) isotropic hole band gap
and  electron band $i=e1,e2$  gaps  $\sim 1 + r_{e} \cos 2\phi $ with $r_{e}>0$; (f) with $r_e<0$. }
\label{fig:FS}
\end{figure}

\section{Three-band model}

We now investigate whether the basic notions which govern the effect of 
disorder on $T_c,T_{nem}$  in the one-band case 
continue to hold in a somewhat more realistic multiband framework appropriate
for the FeSe system that motivated this study, and ask whether a 
disorder-driven $T_c$ enhancement is possible as a consequence of multiband physics.   
Unlike its monolayer and intercalate cousins, bulk FeSe has well-established
Fermi surface hole  pockets at $\Gamma$ and electron pockets at $X$ and $Y$, 
as modeled in Fig. \ref{fig:FS} (a) in the 1-Fe zone.  Although many phenomenological
assumptions for the forms of the interactions on these three bands are possible, we 
assume for simplicity the same Pomeranchuk harmonic for the nematic instability on 
all bands,  leading to the distorted Fermi surface pockets also plotted in the figure.  
We further adopt various simple forms of the BCS pairing interaction on all three 
bands, including pairing anisotropy on one band or the other, as also illustrated 
in Fig. \ref{fig:FS} (b)-({f}).   Because we operate without a microscopic theory, 
we have considerable freedom to choose gap structures consistent with $C_2$ 
symmetry in the nematic phase. We therefore consider anisotropy on the hole and 
the electron pockets independently, although in reality all will distort simultaneously.    
Note that all order parameters on all bands are determined self-consistently. 
Our goal is simply to demonstrate that a $T_c$ enhancement with disorder due 
to competition with nematic order is possible, and deduce what qualitative 
conclusions we can from that novel situation.
\subsection{Electronic structure and pairing interaction}

The Hamiltonian for this three band model is
\begin{eqnarray}
 \mathcal{H} &=& \sum_{\k,i=h,e_1,e_2,\sigma} \left[ (\xi_{\k i}+\Phi_{\k i}  ) c^\dagger_{i,\k,\sigma} c_{i,\k,\sigma}  \right. \nonumber  \\
 &+& \left.  \left ( \Delta_{\k i} c^\dagger_{\k,i,\sigma}c^\dagger_{-\k,i,{-\sigma}} + h.c. \right) \right], \\
 &=& \sum_\k \Psi_\k^\dagger \mathbf{H}_\k \Psi_\k
\label{Eq:Hamiltonian}
\end{eqnarray}
where $c^\dagger_{i,\k,\sigma} (c_{i,\k,\sigma})$ creates (annihilates) a fermion with momentum $\k$ and spin $\sigma$ in the $i^{th}$ band.
 $\Psi$ is an extended Nambu basis for three bands, $\Psi=(c_{-\k,e1\downarrow},c^\dagger_{\k,e1\uparrow},c_{-\k,e2\downarrow},c^\dagger_{\k,e2\uparrow},c_{-\k,h\downarrow},c^\dagger_{\k,h\uparrow})$. The fermionic dispersions in the $C_4$ symmetric phase are,
\begin{eqnarray}
 \xi_h &=& \mu_h - \frac{k^2}{2 m} \\
 \xi_{e1} &=& \frac{k^2_x}{2 m (1+\varepsilon )}+\frac{k^2_y}{2 m (1-\varepsilon )} - \mu_e \\
 \xi_{e2} &=& \frac{k^2_x}{2 m (1-\varepsilon )}+\frac{k^2_y}{2 m (1+\varepsilon )} - \mu_e.
\end{eqnarray}
We assume a circular hole pocket and  slightly elliptic electron pockets with $\varepsilon=0.6$.

The nematic and superconducting order parameters are taken to 
be $\Phi_{\k i}=\Phi_id_{\k i}~ \forall i$,  and $\Delta_{\k i}= \Delta_i{\cal Y}_i(\phi_i)$,
respectively.  The multiband self-consistency equations for the  
nematic and SC orders analogous to Eqs. (\ref{Eq:Phi1band})-(\ref{Eq:Tc1band}) are then obtained as 
\begin{eqnarray}
 \Phi_i &=&  -T \sum_{\omega_n,j,\k_j}  \frac{V^{nem}_{ij} d_{\k_j} \left( i\omega_n - \xi_{\k_j}-\Phi_{\k_j}\right)}{\tilde\omega_n^2+\left( \xi_{\k_j}+\Phi_{\k_j} d_{\k_j}\right)^2}, \label{Eq:Phi3band}\\
 \Delta_{\k_i} &=&- T \sum_{\omega_n,j,\k_j}   \frac{V^{sc}{\cal Y}(\phi_i){\cal Y}(\phi_j)\tilde{\Delta}_{\k_j}}{\tilde\omega_n^2+\left( \xi_{\k_j}+\Phi_{\k_j} d_{\k_j}\right)^2}. \label{Eq:Tc3band}\nonumber
 \end{eqnarray}
Here $\omega_n$ is the fermionic Matsubara frequency at temperature $T$, and
${\cal Y}_i=(1+r_i \cos 2 \phi_i)/\sqrt{1+r^2_i/2}$, where the parameter $r_i$ controls the
anisotropy of the SC order. {\red } Note that while $\phi$ is always measured 
with respect to the positive $x$ axis at each pocket,  $r_{e_i}$ will now be 
assumed to have  same sign on pockets $e_1$ and $e_2$, such that an overall $C_2$ 
state is realized.  Two possible such choices are illustrated in Fig. \ref{fig:FS} (e)-(f).

We use a separable form of interactions for the nematic ($V^{nem}_{ij}d_{\k_i} d_{\k'_j} )$
and the SC order ($V^{sc}_{ij} {\cal Y}_i(\phi_i)Y_j(\phi_j^\prime)$), where summation over the repeated
indices is implied.  
In Eq. (\ref{Eq:Tc3band}), disorder-renormalized quantities are given as multiband generalizations of Eqs. (\ref{Eq:wtild})-(\ref{Eq:deltild_ani}),
\begin{eqnarray}
  i\tilde{\omega}_{i}(\omega_n) &=& i\omega_n -\Sigma_{0i}(\omega_n), \label{Eq:wtild} \\
 {\tilde{\Delta}}^{iso}_i &=& \Delta^{0}_i/\sqrt{1+r_i^2/2} + \Sigma_{1i}(\omega_n) , \label{Eq:deltild}\\
 \Delta^{ani}_i&=& \Delta^0_i r_i/\sqrt{1+r_i^2/2}.
 \label{Eq:deltild_ani}
\end{eqnarray}
Note that the Nambu components of the disorder self-energy
\begin{eqnarray}
\Sigma_{0i}(\omega_n) &=&  n_{imp}  \sum_{j,\k} |u_{ij}|^2 G_{0j}(\k_j,\omega_n) \\
\Sigma_{1i}(\omega_n) &=&   -n_{imp}  \sum_{j,\k} |u_{ij}|^2 G_{1j}(\k_j,\omega_n),
 \label{Eq:Disorder}
\end{eqnarray}
involve both intra- and interband scattering processes via the  impurity scattering potential matrix in the band basis $u_{ij}$, which is taken to have only two elements: $v$ for intra- and $u$ for interband scattering, for all band components, with $\eta\equiv u/v$.   Here $G_{0(1)i}$ is the $\tau_{0(1)}$ component of the $i^{th}$ band's Nambu Green's function and
$\Delta_i^{iso}$ is the isotropic component of the gap function for $i^{th}$ band.
Note that we ignore the $\tau_3$ component of the impurity self-energy, which mainly renormalizes the
chemical potential. 

It is now relatively easy to arrange for nematic order and superconductivity 
to compete, without fine tuning of the interactions.  
Bulk FeSe itself is known to have an order parameter that is highly anisotropic,
with nodes or near-nodes somewhere on the Fermi surface\cite{HirschfeldCRAS}.  
We therefore focus particularly on  cases with large gap anisotropy, either on
the hole pocket (Fig. \ref{fig:FS}(c)-(d)) or on the electron pockets {(Fig. \ref{fig:FS}(e)-(f))}.

Note  we only consider $s_\pm$ type pairing states, with overall sign change
between electron and hole pockets. However, our conclusions are mostly qualitatively
valid for $s_{++}$ states as well, since minimal pairbreaking is required to obtain 
a $T_c$ enhancement in an $s_\pm$ state, and we therefore are forced to assume 
relatively weak interband scattering, such that the sign difference does not play an essential role.

\subsection{Results}
\begin{figure}[h]
\includegraphics[width= 0.95\columnwidth]{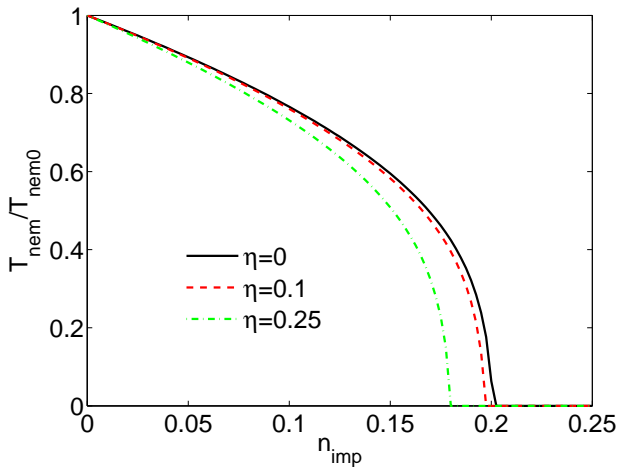}
\caption{Suppression of nematic order by disorder in 3-band model.  }
\label{fig:nem_disorder}
\end{figure}

We first focus on the effect of disorder on the pure nematic state. Both
interband and intraband scattering suppress the nematic order in the current
model, as shown in the inset to Fig. \ref{fig:nem_disorder}, which  displays  the  variation of $T_{nem}$
with increasing impurity concentration. As the interband scattering increases, 
nematicity goes down rapidly; since interband scattering connects Fermi 
surfaces with different signs of $\Phi_\k$, this is equivalent to the effect 
seen for the one-band case in Ref. \onlinecite{Schofield08}.

The relative effects of interband impurity scattering obviously depends now on the gap
structure, i.e. whether a sign changing state is realized.  Evidence for sign changing
gap behavior in this system is limited, and reviewed in Ref. \onlinecite{HirschfeldCRAS}.   
The most significant is probably the recent experiment by   Wang \textit{et. al.}\cite{QWang2016}, 
who observed a ${\bf q}=(\pi,0)$ low-energy inelastic neutron scattering 
resonance very similar to that generally  taken as strong evidence for the $s_\pm$ 
state in other Fe-based superconductors\cite{HKM2011}.
Earlier,  STM\cite{Song2011} and some thermal conductivity and penetration 
depth measurements\cite{Kasahara2014} had reported a gap with nodes.  More recently, 
and possibly on slightly different samples, a gap with very deep minima has been 
suggested to be consistent with STM\cite{Jiao2016}, penetration depth\cite{Teknowijoyo16,broun2016},  
and low temperature thermal conductivity measurements\cite{Hope2016} as well.  
For our purposes, such tiny differences in gap structure are probably irrelevant. 

\begin{figure}[t]
\includegraphics[width= .48\columnwidth]{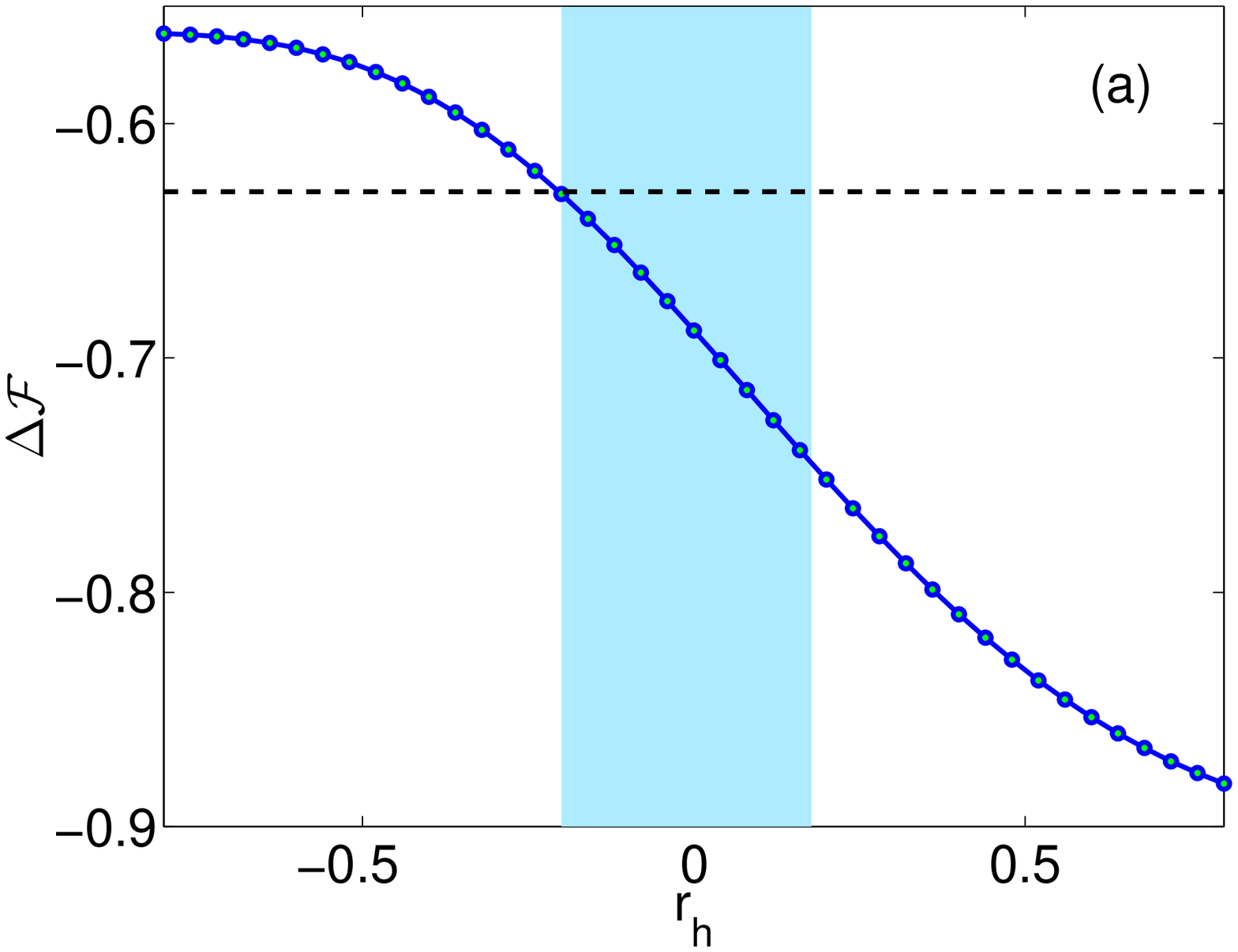}
\includegraphics[width= .48\columnwidth]{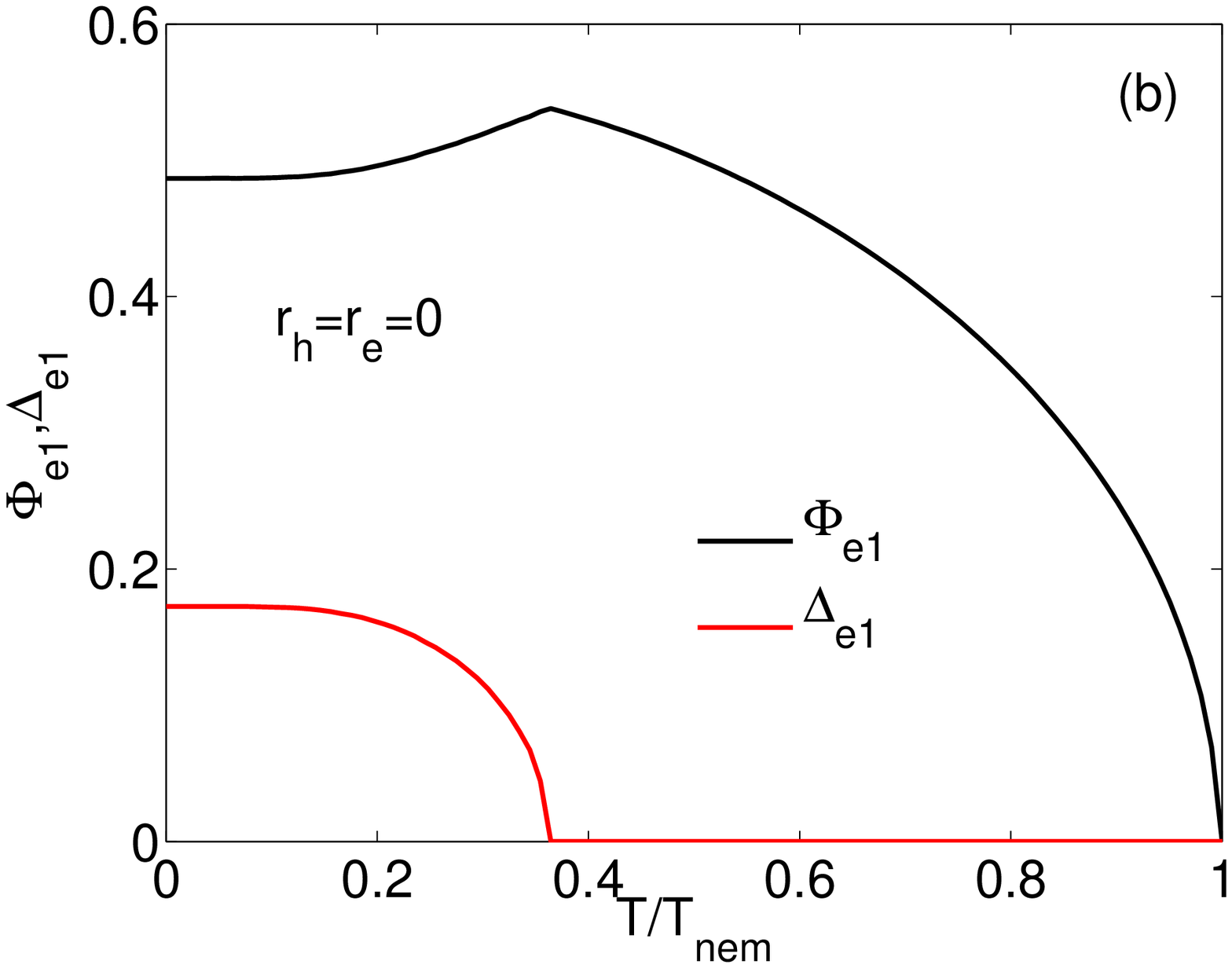}
\includegraphics[width= .48\columnwidth]{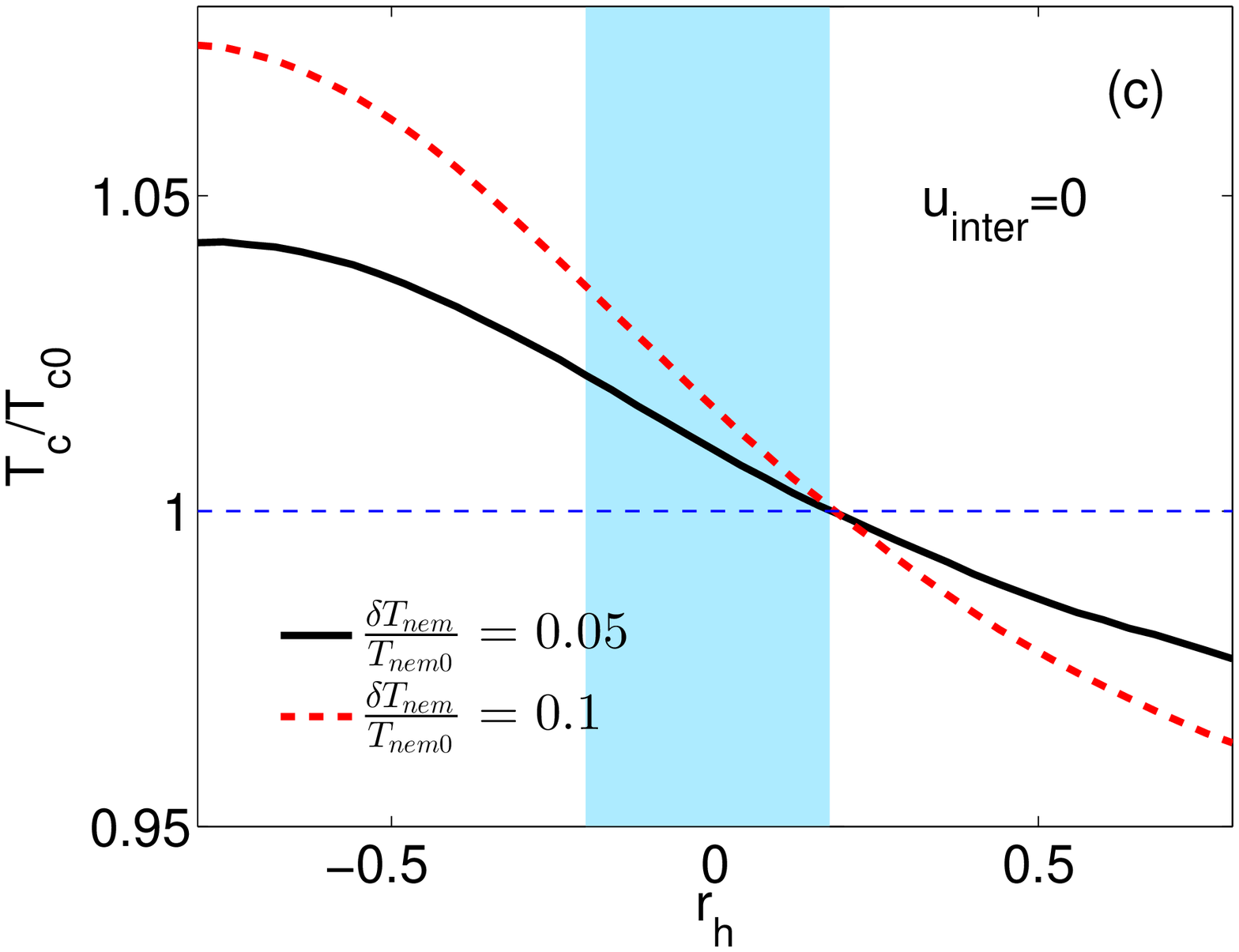}
\includegraphics[width= .48\columnwidth]{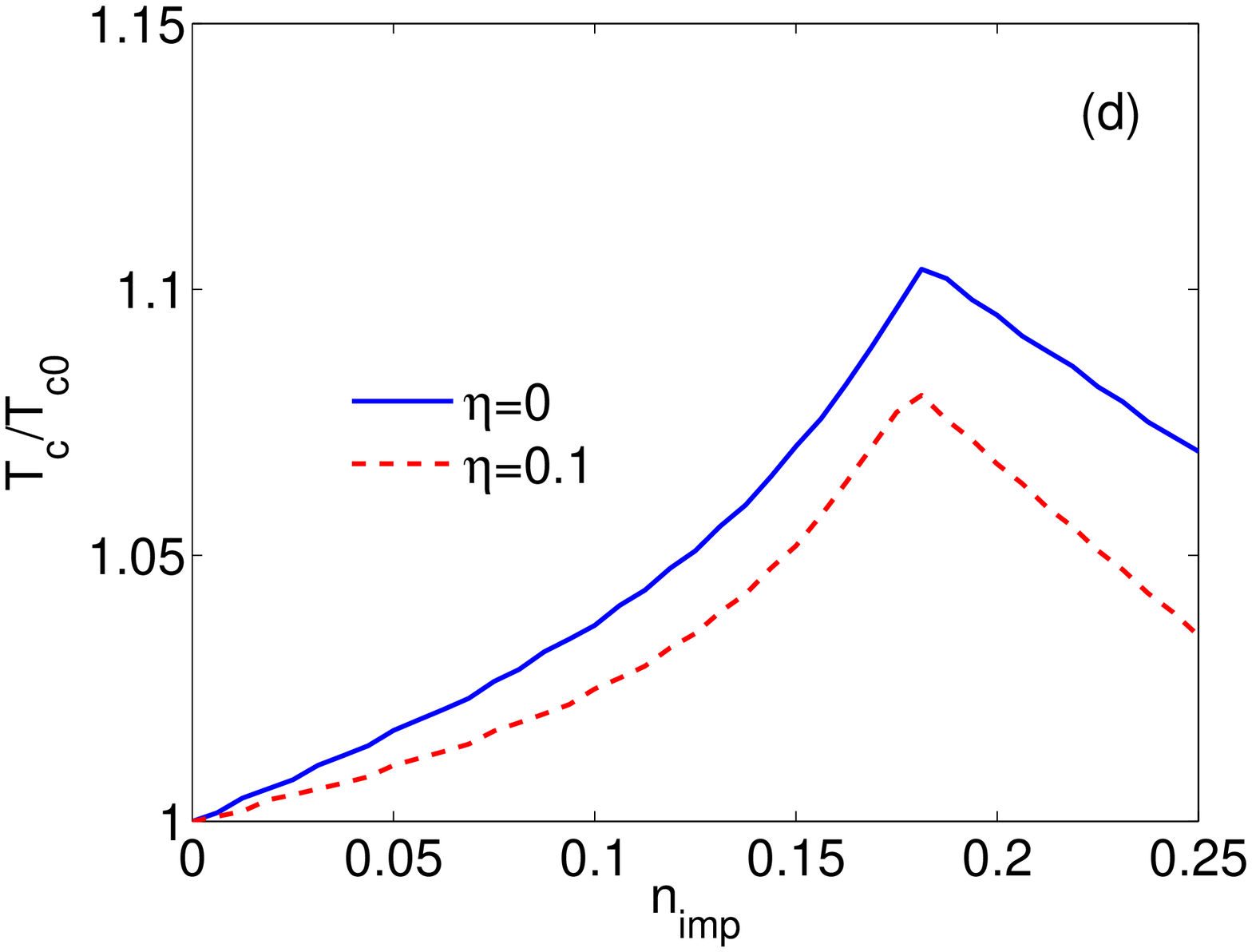}
\caption{Results for 3-band model with anisotropy on hole band only, $\Delta_h \sim (1 + r_h\cos 2\phi)$.  
(a) Free energy of nematic superconductor relative to pure nematic state vs. $r_h$.  (b) Order 
parameters in pure state as function of $T$ for isotropic state.  (c) $T_c$ normalized to $T_{c0}$ 
for pure system vs. anisotropy parameter $r_h$ for two different disorder concentrations, 
corresponding to suppressions of $T_{nem}$ by 5 and 10\%.   Blue shaded region in panels (a) 
and (c) indicates region of thermodynamic stability of coexistence phase.  (d)  $T_c/T_{c0}$ 
vs. disorder scattering rate $\Gamma$
for isotropic $s_\pm$ gap for inter/intraband scattering potential ratio $\eta = u/v$=0.,0.1.    }
\label{fig:holepocketanis_3band}
\end{figure}

The superconducting transition temperature $T_c$ in the presence of disorder and nematicity is now determined
by solving the multiband linearized gap equation, Eq. (\eqref{Eq:Tc3band}),  
which contains the  nematic order parameter $\Phi_{\k i}$, itself determined by Eq. \ref{Eq:Phi3band}.  
We first examine the simplest case,
$r_i=0$ on all bands, i.e. an isotropic gap in the presence of the nematically distorted Fermi surfaces.
The isotropic state in the presence of a single nematic pocket was not 
a stable solution for the nematic distortion chosen.   However, in the presence of the electron
pockets and an interband interaction, an isotropic state in the presence of nematic order can be
lower in energy than the pure nematic one, as seen in Fig. \ref{fig:holepocketanis_3band} (a).  The
competition of the two order parameters is shown explicitly in Fig. \ref{fig:holepocketanis_3band}(b) and
in fact is seen to lead to a modest $T_c$ enhancement with disorder as seen in Fig. \ref{fig:holepocketanis_3band} (c) and (d).   This 
effect may now be enhanced somewhat by considering nonzero $r_h$.  Note that these
effects, while roughly consistent in magnitude with the $T_c$ enhancement effect seen in
experiment, will be suppressed in an $s_\pm$ state by any additional source of pairbreaking, 
e.g. as when the interband scattering rate is increased (Fig. \ref{fig:holepocketanis_3band} (d)).

\begin{figure}[t]
\includegraphics[width= .49\columnwidth]{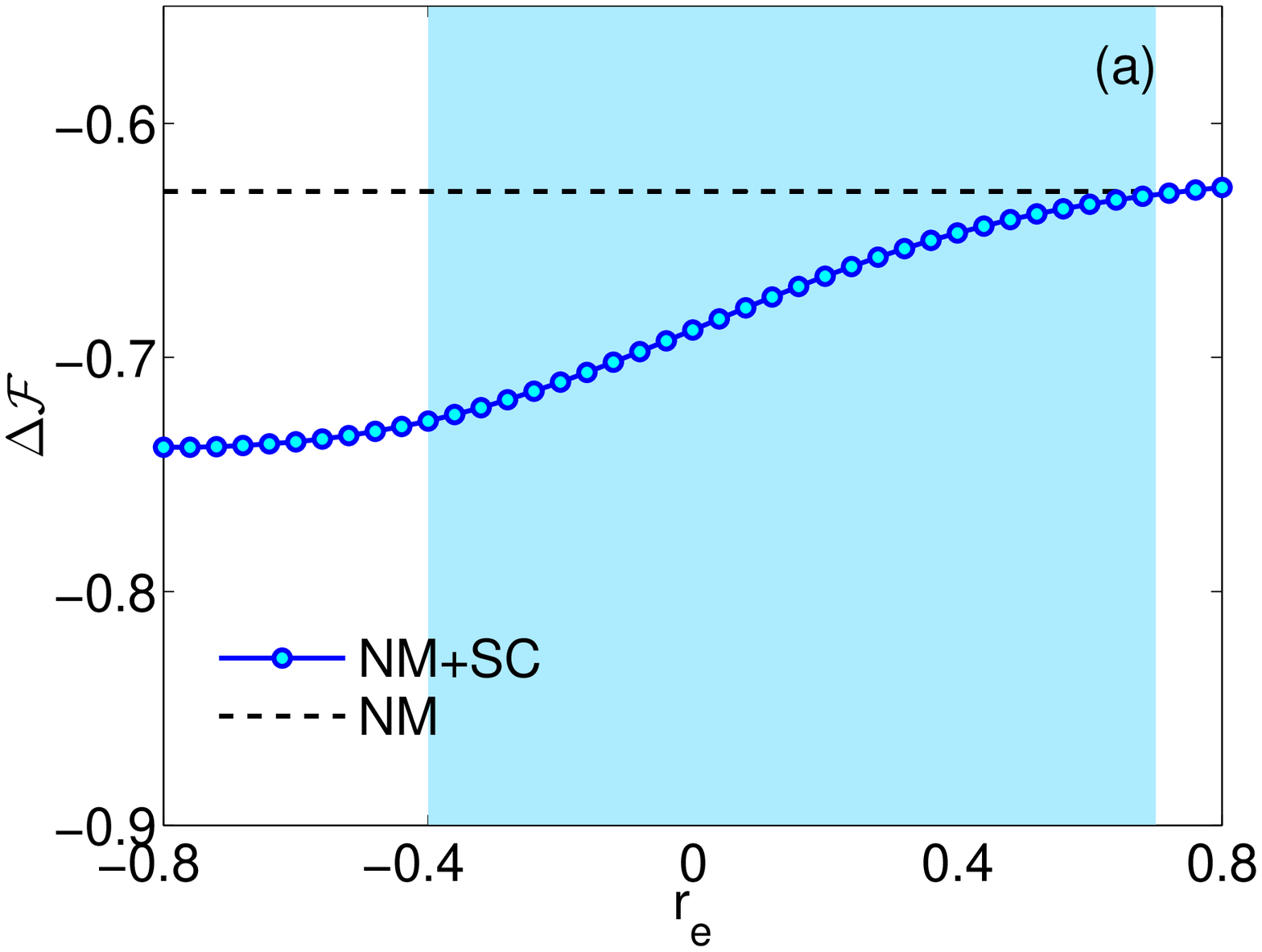}
\includegraphics[width= .47\columnwidth]{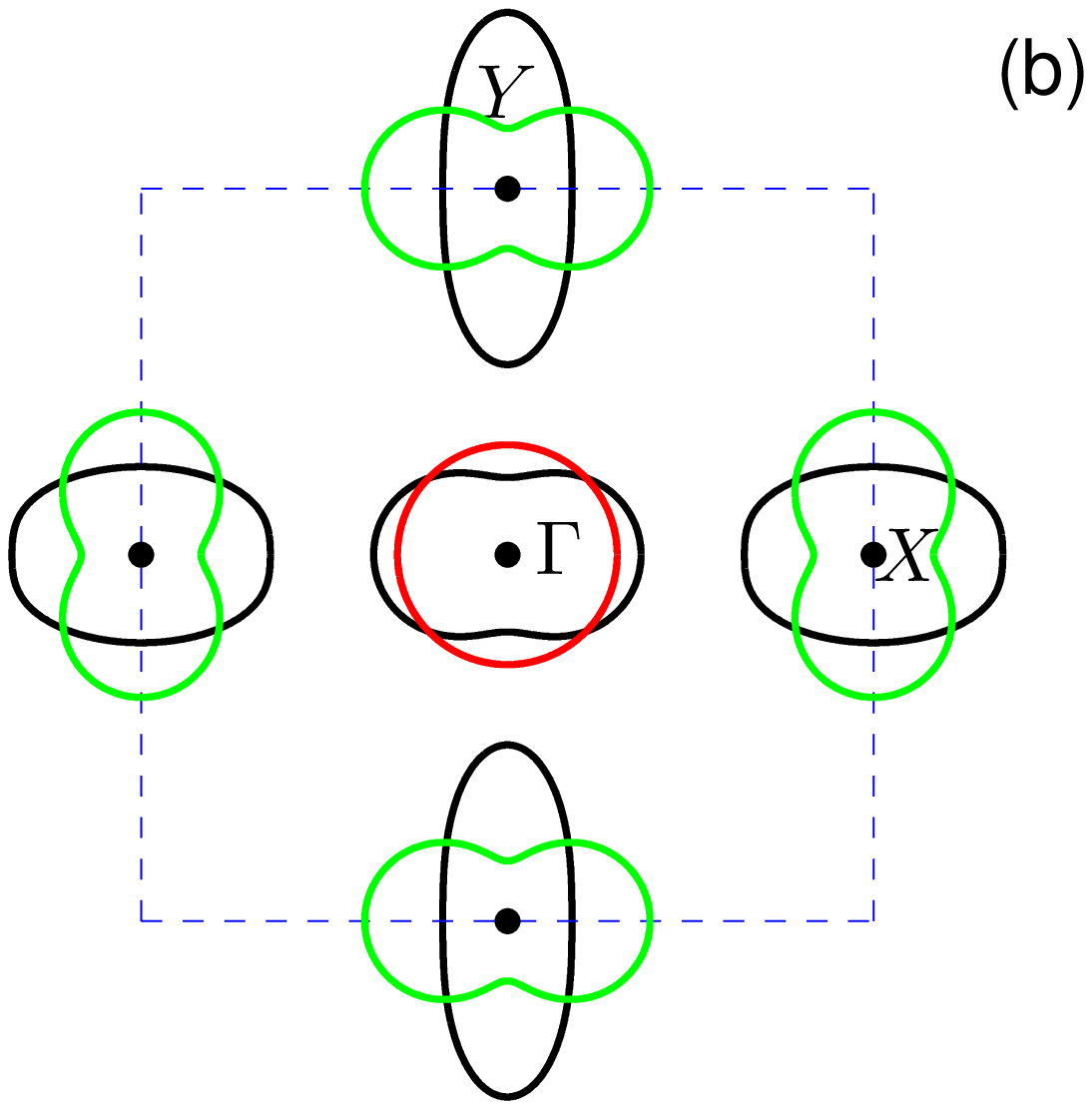}
\includegraphics[width= .49\columnwidth]{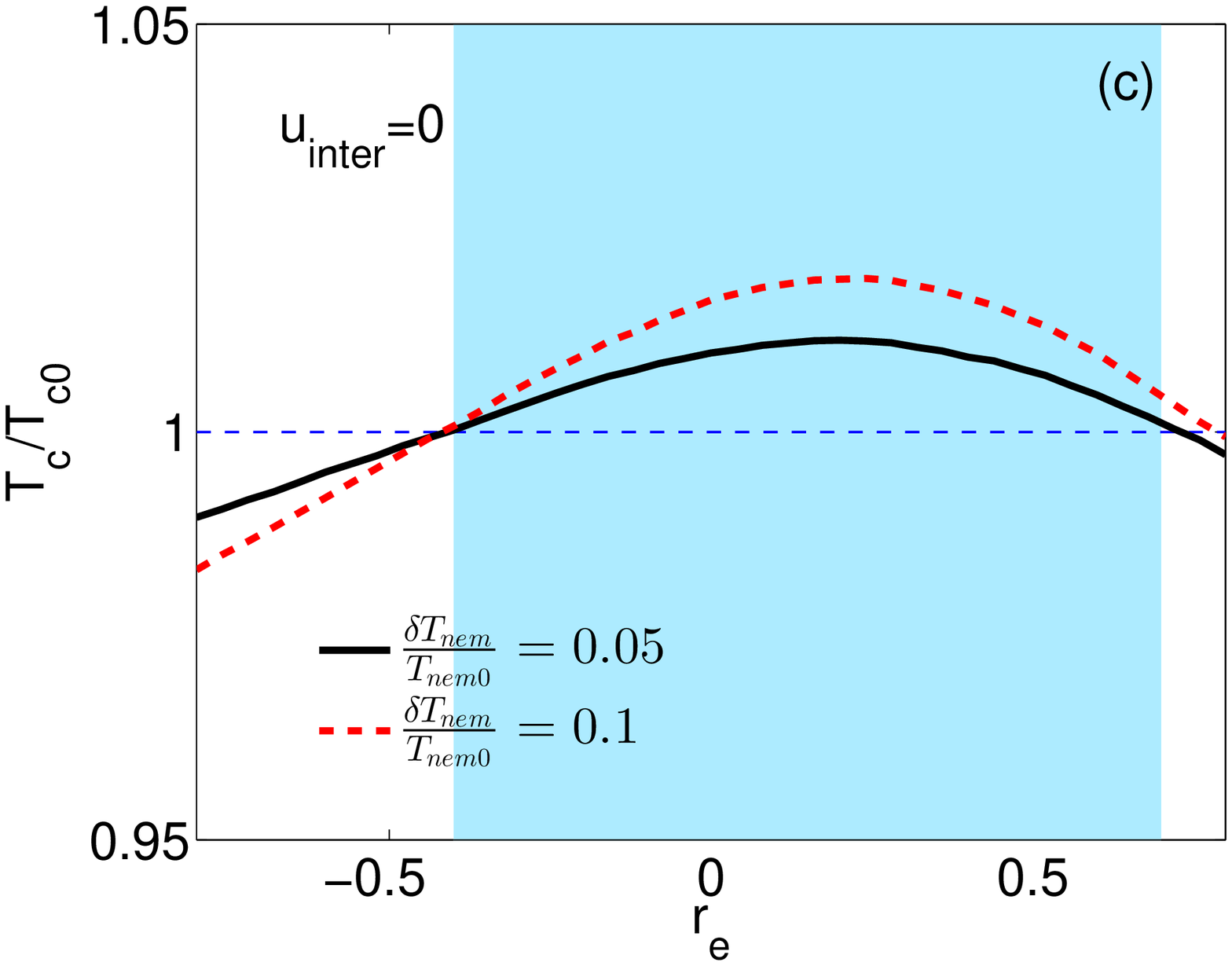}
\includegraphics[width= .47\columnwidth]{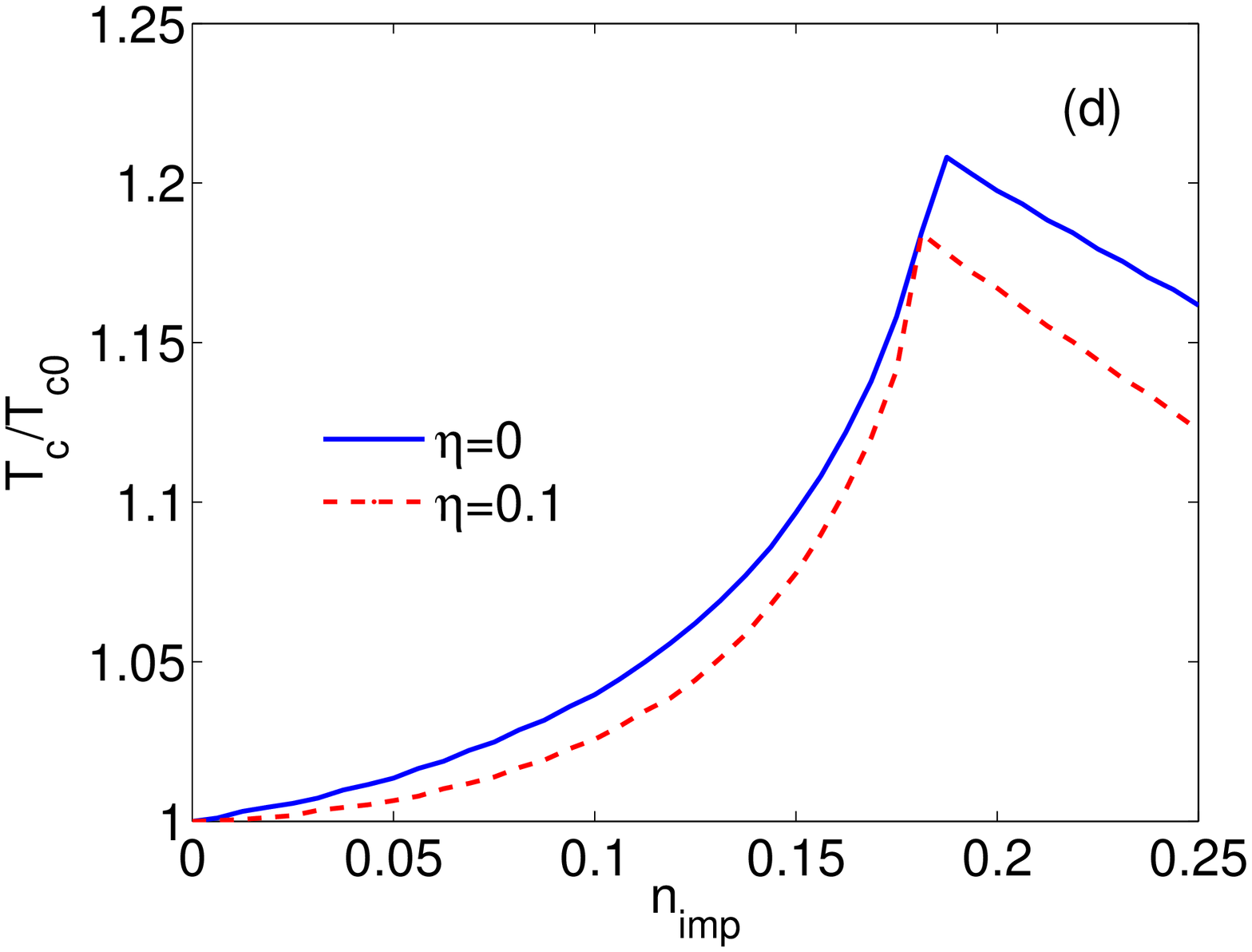}
\caption{  Results for 3-band model with anisotropy on electron band only, $\Delta_h \sim (1 + r_e\cos 2\phi)$.  
(a) Free energy of nematic superconductor relative to pure nematic state vs. $r_e$.  (b) Order 
parameter (green, red line) for $r_h=0,~r_e=0.5$ plotted over nematically distorted Fermi surface.  
(c) $T_c$ normalized to $T_{c0}$ for pure system vs. anisotropy parameter $r_e$ for two different 
disorder concentrations, corresponding to suppressions of $T_{nem}$ by 5 and 10\%.   Blue shaded 
region in panels (a) and (c) indicates region of thermodynamic stability of coexistence 
phase.  (d)  $T_c/T_{c0}$ vs. disorder scattering rate $\Gamma$
for $r_e=0.5$ $s_\pm$ gap for inter/intraband scattering potential ratio $\eta = u/v$=0.,0.1.   }
\label{fig:electron_anis_3band}
\end{figure}

Note that within the current model the ability of the hole band gap anisotropy to
enhance $T_c$ further is limited by the narrow range of stability of this 
state (see Fig. \ref{fig:holepocketanis_3band} (a)-(c)).  It is interesting
therefore to explore the role of gap anisotropy on the electron pockets, which 
we illustrate in Fig. \ref{fig:electron_anis_3band}.  In panels (a) and (c), the 
range of stability of a state which enhances $T_c$ with disorder is found to be 
much wider than in the hole pocket case.  In addition, this range is asymmetric 
with respect to the anisotropy parameter $r_e$, indicating that it is more likely 
to observe $T_c$ enhancement if $r_e>0$, i.e. if the gap minima on the most nematically 
distorted pockets are aligned with the pocket elongation axis (Fig. \ref{fig:FS}).

\section{Discussion}
The motivation for this study has been a recent low-energy electron irradiation experiment by 
Teknowijoyo \textit{et. al.}\cite{Teknowijoyo16}, which found a surprising  enhancement 
of $T_c$ with increasing disorder in FeSe, a system with no long-range magnetic order 
but strong nematic order below the structural transition.  It is worth noting that until
now, all other Fe-based superconductors similarly irradiated  have had their critical
temperatures strongly suppressed by this type of disorder, which should create nearly ideal pointlike potential scatterers. 
This striking result raises several questions about the interplay of superconductivity
and nematicity, and may provide an important clue to the physics of the mysterious FeSe material.   

The experimental situation with regard to competition of superconductivity and
nematic order, required for our  scenario, is currently unclear. In several situations, 
including hydrostatic pressure\cite{medvedev09,kothapalli16} and doping by sulfur 
or field gating\cite{coldea15,Wang_fieldgating16}, $T_s$ decreases while $T_c$ increases, 
implying competition of the two orders. However, B\"ohmer et al. reported seeing no 
direct effect of the onset of superconductivity on the orthorhombic order parameter 
in FeSe crystals\cite{bohmer13}, and recently, Wang et al.\cite{Wang_Meingast2016} 
presented clear evidence that the orthorhombic distortion  was       {\it enhanced} 
by the occurrence of superconductivity in S-doped FeSe crystals. This suggests that 
the interplay between the two orders may be more subtle than assumed here. If our 
approach is qualitatively correct, however, it implies that disorder-induced $T_c$ 
enhancement of the type reported in Ref. \onlinecite{Teknowijoyo16} and discussed 
here should disappear in the S-doped systems.

Clearly one aspect of the simple model presented here is unphysical, namely that
the pairing interaction assumed is ``rigid', in the sense that as the Fermi surface
deformed as $T_c$ is lowered, our model does not capture the concomitant evolution 
of the {\it pairing interaction} before $T_c$ is reached (it does treat the anisotropy 
of the {\it gap} in the presence of a distorted Fermi  surface self-consistently).  
It seems therefore possible  that our model  overestimates competition of 
superconductivity, and nematicity and therefore  artificially favors $T_c$ 
enhancement when nematicity is suppressed.    A more complete microscopic 
treatment of the effect of nematic order on pairing  is outside the scope of 
present manuscript and will be treated
elsewhere.  

\section{Conclusions}
In summary, we have studied the interplay between nematicity and superconductivity for
a system with one band and three bands, modeling the the nematic instability with
a mean field treatment of a $d$-wave Pomeranchuk transition. 
We have shown  that in  several physically plausible circumstances,   nematic order 
competes with superconductivity, and may allow $T_c$ to rise, as observed 
in Ref. \onlinecite{Teknowijoyo16}, when disorder suppresses nematicity.  
Indeed, such an enhancement appears rather natural in the presence of sufficiently 
strong nematicity.  We note that the effect is qualitatively different 
from a $T_c$ enhancement resulting from the competition between 
superconductivity and antiferromagnetism discussed by earlier authors.   

We have further discussed how the $T_c$ enhancement  effect is  sensitive to
the degree and orientation of the gap anisotropy with respect to the deformed Fermi surface.  
In particular, we showed that superconducting gaps with minima along the stretched axis 
of the deformed Fermi surface are easily suppressed by nematic order. Upon introduction
of disorder, which rapidly destroys nematic order,  such states show 
significant $T_c$ enhancement.  In the one-band case, these states do not 
appear to be thermodynamically stable, but they are stabilized by the addition 
of additional pockets, as in the three-band model studied here.    In contrast, 
SC states with gap maxima along the stretched direction of Fermi surface, do not 
compete strongly with nematic order,   so $T_c$  of an $s_\pm$ state is suppressed as usual by disorder. 
\vskip .2cm

\emph{Acknowledgments.}  The authors would like to dedicate this paper to Florence Rullier-
Albenque, whose pioneering studies of electron irradiation disorder in cor-
related electron systems inspired them. The authors are grateful for useful
discussions with B M Andersen, M H Christensen, I Paul and R Prozorov.  
PJH was supported by NSF-DMR-1407502. VM was supported by the Laboratory Directed
Research and Development Program of Oak Ridge National Laboratory, managed by
UT-Battelle, LLC, for the U. S. Department of Energy. 

\appendix
\section{Free Energy}
Here we calculate the free energy of our system including  mean field 
treatments of nematic and superconducting order. In general, the free energy is given by
\begin{eqnarray}
\mathcal{F}= \langle \mathcal{H} \rangle - T \mathbf{S}
\label{eq:freeE1}
\end{eqnarray}
where the entropy $\mathbf{S}$ is
\begin{eqnarray}
\mathbf{S}= -2\sum_{k} \left[ f(E_k) \log E_k + (1-f(E_k)) \log (1-f(E_k)) \right].\nonumber\\
\end{eqnarray}
Here $E_k=\sqrt{\left( \xi_k +\Phi_k \right)^2+ \Delta^2_k}$, $f$ is the Fermi function. 
The first term in Eq. \eqref{eq:freeE1} is the expectation value of the mean field 
Hamiltonian, which can be evaluated using coherence factors 
\begin{eqnarray}
u^2_k &=& \frac{1}{2} \left[ 1+ \frac{\xi_k + \Phi_k}{E_k} \right] \\
v^2_k &=& \frac{1}{2} \left[ 1 - \frac{\xi_k + \Phi_k}{E_k} \right].
\end{eqnarray}
For the one band case, the expectation value of the kinetic energy may be written as,
\begin{eqnarray}
\langle \mathcal{H}_{kin} \rangle &=&\langle \sum_{k,\sigma}  \tilde{\xi}_k c^{\dagger}_{k,\sigma} c_{k,\sigma} \rangle \\
&=&  \sum_{k,\sigma}  \left(1 - \tanh\left[\frac{E_k}{2T}\right] \frac{\tilde{\xi}_k}{E_k}\right) \tilde{\xi}_k,
\end{eqnarray}
where $\tilde{\xi_k}=\xi_k+\Phi_k$. The potential energy term for the one band case is
\begin{eqnarray}
\langle \mathcal{H}_{pot} \rangle &=& -\frac{\Phi_0^2}{V_{nem}} - \frac{\Delta^2_0}{|V_{sc}|},
\end{eqnarray}
and for three bands,
\begin{eqnarray}
\langle \mathcal{H}_{pot} \rangle &=& -\frac{2\phi_h \left(\Phi_{e1}+\Phi_{e2} \right)}{V_{nem}} + \frac{\Delta_h \left(\Delta_{e1}+\Delta_{e2}\right)}{|V_{sc}|},
\end{eqnarray}
Here we have assumed only interband interactions for both orders. 

Note that expressions for $\cal{F}$ in both one- and three-band cases manifestly
reduce to the correct expressions in either the pure nematic or superconducting states.
\end{document}